\documentclass[11pt,twoside]{article}
\usepackage{./asp2014}

\aspSuppressVolSlug
\resetcounters

\begin{document}

\title{How can astrotourism serve the sustainable development goals? The Namibian example.}
\author{Hannah~Dalgleish$^{1,2}$, Getachew~Mengistie$^1$, Michael~Backes$^{1,3}$, Garret~Cotter$^2$, and Eli~Kasai$^1$
\affil{$^1$Dept. of Physics, University of Namibia, Pionierspark, Windhoek, Namibia;}
\affil{$^2$Dept. of Physics, University of Oxford, Keble Rd, Oxford, OX1 3RH, UK; \email{hannah.dalgleish@physics.ox.ac.uk}}
\affil{$^3$Centre for Space Research, North-West University, Potchefstroom, South Africa}}

% This section is for ADS Processing.  There must be one line per author.
\paperauthor{Hannah~Dalgleish}{hannah.dalgleish@physics.ox.ac.uk}{0000-0002-8970-3065}{University of Namibia}{Physics}{Windhoek}{Khomas}{}{Namibia}
\paperauthor{Getachew~Mengistie}{ghionboy@gmail.com}{}{University of Namibia}{Physics}{Windhoek}{Khomas}{}{Namibia}
\paperauthor{Michael~Backes}{mbackes@unam.na}{0000-0002-9326-6400}{University of Namibia}{Physics}{Windhoek}{Khomas}{}{Namibia}
\paperauthor{Garret~Cotter}{garret.cotter@physics.ox.ac.uk}{}{University of Oxford}{Physics}{Oxford}{Oxfordshire}{OX1 3RH}{UK}
\paperauthor{Eli~Kasai}{ekasai@unam.na}{}{University of Namibia}{Physics}{Windhoek}{Khomas}{}{Namibia}

\begin{abstract}
Astrotourism brings new opportunities to generate sustainable socio-economic development, preserve cultural heritage, and inspire and educate the citizens of the globe. This form of tourism can involve many different activities, such as visiting observatories or travelling to remote areas to experience an evening under a pristine, dark night sky. Together, our UK-Namibian collaboration is working to develop and showcase astrotourism in Namibia, and to enhance the possibility for astrotourism worldwide.
\end{abstract}

\section{Introduction}

Since time immemorial, the night skies have been a source of inspiration for humankind. People have been influenced by the stars and their arrangements on the sky, and have studied the positions of the stars for agriculture, timekeeping, and navigation --- some indigenous communities continue to use this knowledge today. Yet, in modern times, the majority of celestial objects are obscured by light pollution; more than 80\% of the world’s population live under light-polluted skies (\citealt{Falchi16}). This has led to an increased desire to seek out the dark, and reconnect with our cultural heritage.

\section{Astrotourism for development}

Astronomy and sustainable development go hand in hand, as several examples on the African continent have already shown (\citealt{Povic18,McBride18,Strubbe21}). Where development through astronomy-related capacity-building projects has received notable attention, new possibilities arise through travel and tourism. 

As far as we are aware, the potential for astrotourism and development was first realised in South Africa (\citealt{Govender09}). The construction of SALT (Southern African Large Telescope)\footnote{\url{https://www.salt.ac.za/}} in the Northern Cape attracted tourists to the area, which in turn stimulated the creation of jobs and small businesses. Since then, astrotourism has been identified as one of the Flagships of the International Astronomical Union's Office of Astronomy for Development (IAU OAD)\footnote{\url{http://www.astro4dev.org/flagship-projects/}}. 
The benefits of astrotourism can also be understood through the United Nations' Sustainable Development Goals (SDGs)\footnote{\url{https://sdgs.un.org/}}. This includes, but is by no means limited to, Goals 4, 5, 8, 9, 10, and 11: quality education; gender equality; decent work and economic growth; industry, innovation, and infrastructure; reduced inequalities; and sustainable communities (\citealt{Dalgleish20}). 

Where astrotourism is often centred around observatories, it is also possible to attract tourists to rural, dark sky areas which do not have any significant astronomical infrastructure. Dark sky tourism, a branch of astrotourism, focuses exclusively on activities which require dark skies, and can therefore bring significant advantages to rural communities (\citealt{Dalgleish21a}). Added benefits are generated through the conservation of the night sky, via organisations like the International Dark-Sky Association (IDA)\footnote{\url{https://www.darksky.org/}}. Communities can apply to the IDA for dark sky status, enhancing the visibility and protection of dark sky oases, and fostering increased tourism and local economic activity (\citealt{Mitchell19}). There are currently more than 130 certified International Dark Sky Places in the world, although only two of these are in Africa.

\section{Astrotourism in Namibia}

Namibia is ideal for astronomy-related tourism experiences. The country has a very low population density (3 people/km$^2$), and so there is little-to-no light pollution in the vast majority of the country. Namibian weather is an additional advantage --- dry conditions make for fewer clouds and clearer conditions to observe celestial objects. Alongside the amazing skies, the country is developing its capacity for astrophysics research (\citealt{Backes18}) and has many exciting astronomy-related sites to offer, including the Hoba meteorite (the largest meteorite in the world); the NamibRand Nature Reserve\footnote{\url{http://www.namibrand.com/dark-sky.html}} (Africa's first International Dark Sky Reserve); and the world-leading High Energy Spectroscopic System (H.E.S.S.) observatory (\citealt{deNaurois18}).

Tourism is one of Namibia's key industries, accounting for 10.3\% of Gross National Product in 2019 (\citealt{wttc20}). Despite being one of the best places in the world to see unpolluted night skies, Namibian tourism typically revolves around safaris, hiking, indigenous culture, and trophy hunting. There is, however, a small community of amateur astronomers who have been visiting ``astrofarms'' over the past few decades to do astrophotography. 

Namibia is also at the forefront of community conservation; $\sim$20\% of Namibia is covered by 86 communal conservancies --- self-governing, democratic entities dedicated to the management and conservation of natural resources within social, cultural and economic contexts. Although not explicitly recognised, parallels can be drawn between these conservancies and the ethos of dark sky preservation. A recent meeting with the Ju/’Hoansi Traditional Authority (TA) in the Nyae Nyae conservancy, showed that astrotourism was unknown to them, but they were eager to learn more. Tourism in the area has essentially come to a standstill (due to the Covid-19 pandemic) and so astrotourism presents an opportunity to reignite tourism activity in post-Covid times. The TA also shared that much of their indigenous starlore has been lost, and so astrotourism provides further opportunities to help recover and retain this endangered knowledge.

One of the main hurdles in implementing astrotourism is the training of tour guides in astronomy and stargazing. A tour guide qualification on this topic already exists under the Namibia Training Authority, but it has been mostly unobtainable given a lack of accessible training, and is also in need of review. We plan to update the qualification, as well as devise an astronomy course for tour guides (\citealt{Dalgleish21b}) --- we hope to deliver some of this training ourselves, as well as to `train the trainer', to ensure sustainability. Overall, there is great scope for Namibia to expand and diversify its tourist demographic via astrotourism, and with little cost (\citealt{Stone19}). 

\section{Summary}

Astrotourism makes use of unpolluted nightscapes as a natural and infinite resource. As dark skies become more scarce, many countries in the Global South have a unique opportunity to offer their dark skies to travellers who seek to reconnect with the heavens. Astronomy activities can also offer a sustainable and meaningful tourism experience, while keeping in line with the conservation of natural resources and cultural heritage. Moreover, it can be implemented at minimal cost, and with minimal infrastructure. In all, astrotourism presents opportunities for development through education, sustainable job creation, reduced inequalities, and the preservation of indigenous knowledge. For these reasons, astrotourism fulfils several of the SDGs and shows great potential for sustainable development in countries like Namibia.

\acknowledgements We acknowledge support from the UKRI STFC Global Challenges Research Fund project ST/S002952/1 and Exeter College, Oxford.

%\bibliography{editor}  % For BibTex

% For non-BibTex:

\end{document}